\documentstyle[twoside,fleqn,epsfig,espcrc2]{article}

\newcommand{\AmS}{{\protect\the\textfont2
  A\kern-.1667em\lower.5ex\hbox{M}\kern-.125emS}}

\title{Using {\sc top-c} for Commodity Parallel Computing in Cosmic Ray 
Physics Simulations}
\author{G. Alverson\address{Department of Physics, Northeastern  
        University, Boston, MA 02115, 
        USA}\thanks{george.alverson@cern.ch},
        L. A. Anchordoqui$^a$\thanks{doqui@hepmail.physics.neu.edu}, 
        G. Cooperman\address{College of Computer Science, Northeastern 
        University, Boston, MA 02115, USA}\thanks{gene@ccs.neu.edu},
        V. Grinberg$^b$\thanks{victor@ccs.neu.edu},
        T. P. McCauley$^a$\thanks{mccauley@hepmail.physics.neu.edu}, 
        T. Paul$^a$\thanks{tom.paul@hepmail.physics.neu.edu},
        S. Reucroft$^a$\thanks{stephen.reucroft@cern.ch}, $\,$ and 
        J. D. Swain$^a$\thanks{john.swain@cern.ch}} 
\begin{document}

\begin{abstract}

{\sc top-c} (Task Oriented Parallel~C) is a freely available package for
parallel computing.  It is designed to be easy to learn and to have
good tolerance for the high latencies that are common in commodity
networks of computers.  It has been successfully used in a wide range
of examples, providing linear speedup with the number of computers.  A
brief overview of {\sc top-c} is provided, along with recent experience with
cosmic ray physics simulations.

\end{abstract}
\maketitle

\section{Introduction}

Ultra high energy cosmic rays are observed indirectly through 
detection of the extensive air showers that are produced when
they travel through the atmosphere. 
To adequately interpret the measured observables and to be able to infer the 
properties of the incident primary particle, a full Monte Carlo treatment of 
the extensive air shower is neeeded. The CPU time required rises with 
the primary energy. For example, for primary energies around  $10^{20}$~eV 
a shower contains about $10^{11}$ secondary particles.
The amount of computing time required to follow all the particles seems 
to be prohibitive. Traditionally, sampling techniques are 
used to reduce the number of particles tracked \cite{hillas}.

In this article we describe an ongoing program to use 
commodity parallel computing for fast 
Monte Carlo simulations \cite{neu6}. The aim is to go beyond the simple 
event-level parallelism which is commonly used today and actually run 
individual events faster than would be possible on a single workstation or PC.

\section{ {\sc geant4} }
    
For a variety of reasons, in no
small part driven by the wish to work with software which is likely
to see use in the future, we decided to try to parallelize {\sc geant4}
\cite{geant}, 
the C++ rewrite of the older ({\sc fortran77}) {\sc geant3}. 
{\sc geant4} is an object-oriented simulation package that provides general-purpose
tools for defining and simulating detector geometry, material properties,
particle transport and interactions, visualization, and all relevant
physics processes. Its versatility allows it to be employed in applications beyond
its traditional usage in High Energy Physics experiments, from the medical and
biological sciences to Cosmic Ray Physics \cite{us}.    
 
\section{{\sc top-c}}

{\sc top-c} (Task Oriented Parallel~C) \cite{topc} was initially designed
with two goals in mind:
\begin{enumerate}
\item to provide a framework for easily developing parallel applications;
\item to build in the ability to tolerate the high latency typically found
on Beowulf clusters.~\footnote{The term ``Beowulf cluster'' refers to a 
cluster of systems running Linux and connected by ethernet.}
\end{enumerate}
The package is freely available~\cite{gene_web}.
The same application
source code has been run under shared and distributed memory (SMP, IBM
SP-2, NoW, Beowulf cluster).  A sequential {\sc top-c} library is also provided to
ease debugging.  The largest test to date was a
computer construction of Janko's group over three months using
approximately 100 nodes of an IBM SP-2 parallel computer at Cornell
University \cite{TOPC-SP2}.

The {\sc top-c} programmer's model \cite{topc} is a master-slave
architecture based on three key concepts:
\begin{enumerate}
\item {\it tasks} in the context of a master/slave architecture;
\item global {\it shared data} with lazy updates; and
\item {\it actions} to be taken after each task.
\end{enumerate}
Task descriptions (task inputs) are generated on the master, and
assigned to a slave.  The slave executes the task and returns the
result to the master. The master may update shared data on all
processes. Such global updates take place on each slave after the
slave completes its current task.  The programmer's model for {\sc top-c} is
graphically described below.

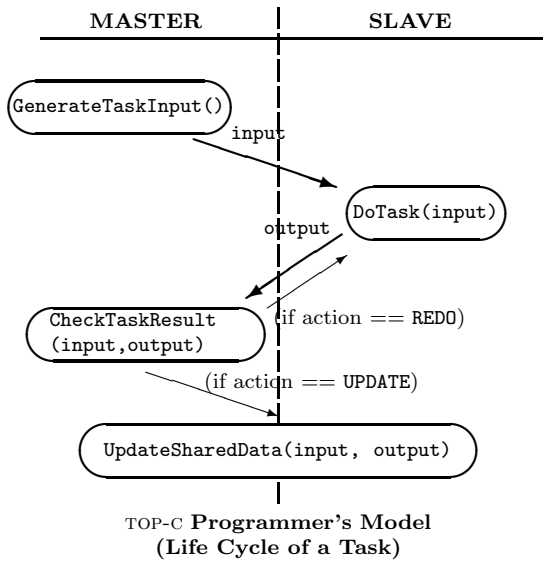
\begin{figure}[htb]\label{topc-fig}
{\footnotesize
\begin{center}\label{diagram}
\setlength{\unitlength}{2pt} 
\begin{picture}(100,95)
\put(25,90){\makebox(0,0)[b]{\bf MASTER}}
\put(75,90){\makebox(0,0)[b]{\bf SLAVE}}
\put(50,00){\line(0,1){4}}
\multiput(50,15)(0,5){16}{\line(0,1){4}} 
\thicklines \put(5,88){\line(1,0){95}}
\put(20,75){\oval(42,10)}  
\put(20,75){\makebox(0,0){\tt GenerateTaskInput()}}
\put(78,55){\oval(30,10)} 
\put(78,55){\makebox(0,0){\tt DoTask(input)}}
\put(25,32){\oval(45,10)} 
\put(38,32){\makebox(0,0){\parbox[t]{125pt}{\tt
CheckTaskResult \hbox{\ \ \ \ \ \ \ \ \ \ \ \ } (input,output)}} }
\put(50,10){\oval(75,10)}
\put(50,10){\makebox(0,0){\tt UpdateSharedData(input, output)}}
\thinlines
\thicklines
\put(34,69){\vector(3,-1){27}} 
\put(41,69){\makebox(0,0)[bl]{\tt input}}
\put(62,51){\vector(-3,-2){18}} 
\put(60,51){\makebox(0,0)[br]{\tt output}} \thinlines
\put(48,37){\vector(3,2){15}} 
\put(49,37){\makebox(0,0)[tl]{(if action == {\tt REDO})}}
\put(25,25){\vector(3,-1){25}} 
\put(36,22){\makebox(0,0)[bl]{(if action == {\tt UPDATE})}}
\end{picture}\break
  \hbox{{\sc top-c} {\bf  Programmer's Model}}\break
   \hbox{\bf (Life Cycle of a Task)}
\end{center}
}
\end{figure}

\section{ Parallelization of {\sc geant4} Using {\sc top-c} }

The task-oriented approach of {\sc top-c} is ideally suited to parallelizing
legacy applications. The tactic in parallelizing {\sc geant4} was to perturb
the existing software as little
as possible and to modify just the section of the code which handles
particle tracking and interaction (a frequent operation) to allow it to
run on multiple CPU's. The largest difficulty was in {\it
marshalling} and {\it unmarshalling} the C++ {\sc geant4} track 
objects that had
to be passed to the slave processes.  Marshalling is the process by
which one produces a representation of an object in a contiguous
buffer suitable for transfer over a network, and unmarshalling is the
inverse process.

We developed a 6-step software methodology to
incrementally parallelize {\sc geant4}, allowing us to isolate
individual issues. The six steps were:
\begin{enumerate}
\item the use of \*.icc (include) files to isolate the code
from the original {\sc geant4} code;
\item collecting the code of the inner loop in a separate
routine, {\tt DoTask()}, whose input was a primary particle track,
and whose output was the primary and its secondary particles;
\item marshalling and unmarshalling the C++ objects for particle tracks
\item integrating the marshalled versions of the particle tracks
with the calls to {\tt DoTask()};
\item adding calls to {\sc top-c} routines such as {\tt TOPC\_init()}, 
{\tt TOPC\_submit\_task\_input} and then testing as the marshalled
particle tracks were sent across the network;
\item and finally adding {\tt CheckTaskResult()}, which inspected
the task output, and added the secondary tracks to the stack, for
later processing by other slave processes.
\end{enumerate}

Prior to the fifth step, all debugging was in a sequential setting.
The maturity of the {\sc top-c} library then allowed us to create fully
functioning parallel code in less than a day.

\section{Discussion }

{\sc geant4} (approximately 100,000 lines of C++ code) was successfully
parallelized using {\sc top-c}.  In the future we plan to perform
timing tests on a long run using many processors.  Initial results
for the example described
indicate that a single task in our application requires approximately 1~ms
of CPU time.  Hence, it will be essential to submit approximately 100
particles for a single slave process to compute in order to overcome
network overhead.  Optimization of the parallel implementaion is
underway.

{\sc top-c} seems to be well-suited to the problem
of parallelizing {\sc geant4}, and would likely be well-suited to other
high energy physics and cosmic ray applications as well. Its flexibility and
simplicity makes it possible to envision enormous speedups for {\sc geant4}
within a single event, something not often considered in high energy
experiments, but offering advantages over the usual 
event-by-event parallelism, especially during interactive data
analysis and code or hardware design.  

Of particular interest is the parallelization of existing cosmic
ray simulation programs such as {\sc aires} \cite{sergio} and {\sc corsika} 
\cite{corsika}.
Although written in {\sc fortran}, such programs are in fact often converted
to C for compilation using f2c, and can certainly be linked with other
C programs, so we anticipate no major obstacles.
We are always 
interested in collaboration with other
groups who may have needs for the speedups that our methodology
offers.


\begin{thebibliography}{99}


\bibitem{hillas} A. M. Hillas, in {\it Proc. of the 16$^{\rm th}$
International Cosmic Ray Conference}, Tokyo, Japan, 1979
(University of Tokyo, Tokyo, 1979), Vol.8,p.7.; updated in,
{\it Proc. of the 17$^{\rm th}$ International Cosmic Ray Conference},
Paris, France, 1981 (CEN, Saclay, 1981), Vol.8,p.183.
\bibitem{neu6} The first analyses were discussed in, 
G. Cooperman, L. Anchordoqui, V. Grinberg, 
T. McCauley, S. Reucroft and J. Swain, {\it Scalable Parallel 
Implementation of Geant4 Using Commodity Hardware and Task 
Oriented Parallel C}, in Proc. CHEP 2000 (only available in the 
CD version) [hep-ph/0001144].
\bibitem{geant} http://wwwinfo.cern.ch/asd/geant/
\bibitem{us} L. A. Anchordoqui {\it et al.}, [astro-ph/0006141] 
and [astro-ph/0006142]. 
\bibitem{topc} G.~Cooperman, ``TOP-C:  A Task-Oriented Parallel~C
Interface'', {\sl $5^{\hbox{th}}$ International Symposium on High
Performance Distributed Computing} (HPDC-5), IEEE Press,  1996,
pp.~141--150.
\bibitem{gene_web} ftp://ftp.ccs.neu.edu/pub/people/gene/topc  
\bibitem{TOPC-SP2} G.~Cooperman, W.~Lempken, G.~Michler and M.~Weller,
``A New Existence Proof of Janko's Simple Group $J_4$'',  {\sl
Progress In Mathematics}~{\bf 173}, Birkhauser, 1999, pp.~161--175.
\bibitem{sergio} S. Sciutto, 
{\it Air Shower Simulations with the} {\sc aires} {\it system},
in {\it Proc. XXVI International Cosmic Ray Conference}, (Eds. D.
Kieda, M. Salamon, and B. Dingus, Salt Lake City, Utah, 1999)
vol.1, p.411, [astro-ph/9905185] at http://xxx.lanl.gov.
\bibitem{corsika} D. Heck {\it et al.}, {\sc corsika}  
{\it (COsmic Ray Simulation for KASCADE)}, FZKA6019 (Forschungszentrum 
Karlsruhe) 1998; updated by D. Heck
and J. Knapp, FZKA6097 (Forschungszentrum Karlsruhe) 1998.
\end{thebibliography}
\end{document}